\documentstyle[12pt]{article}
\begin{document}
\begin{titlepage}
\renewcommand{\thefootnote}{\fnsymbol{footnote}}

\begin{center} \Large
{\bf Theoretical Physics Institute}\\
{\bf University of Minnesota}
\end{center}
\begin{flushright}
TPI-MINN-94/25-T\\
UMN-TH-1263-94\\
UND-HEP-94-BIG07\\
OUT-4102-51\\
July, 1994
\end{flushright}
\vspace{.3cm}
\begin{center} \Large
{\bf How to Measure  Kinetic Energy of the Heavy
\\
Quark Inside B Mesons?}
\end{center}
\vspace*{.3cm}
\begin{center} {\Large
I. Bigi $^{a,b}$, A.G. Grozin $^{c,d}$,
M. Shifman $^e$, N.G. Uraltsev $^{a,f}$,
A.~Vainshtein $^{e,d}$} \\
\vspace{.4cm}
{\normalsize $^a${\it TH Division, CERN, CH-1211 Geneva 23,
Switzerland}\footnote{During the academic year 1993/94}\\
$^b${\it Dept. of Physics,
Univ. of Notre Dame du
Lac, Notre Dame, IN 46556, U.S.A.}\footnote{Permanent address}\\
$^c$ {\it Physics Department, Open University, Milton Keynes MK7
6AA,
UK}\\
$^d$ {\it Budker Institute of Nuclear Physics, Novosibirsk
630090, Russia}\\
$^e$ {\it  Theoretical Physics Institute, Univ. of Minnesota,
Minneapolis, MN 55455}\\
$^f$ {\it St. Petersburg Nuclear Physics Institute,
Gatchina, St. Petersburg 188350, Russia}$^\dagger$\\
\vspace{.3cm}
e-mail addresses:\\
{\it bigi@cernvm.cern.ch, a.grozin@open.ac.uk,
shifman@vx.cis.umn.edu,
vainshte@vx.cis.umn.edu}\\
\vspace*{.4cm}
}
{\Large{\bf Abstract}}
\end{center}

\vspace*{.2cm}
We discuss how one can determine the average kinetic
energy of the heavy quark inside heavy mesons from
 differential distributions in $B$ decays. A new, so-called
third, sum rule for
the $b\rightarrow c$ transition is derived in the small velocity (SV)
limit. Using this sum rule and the existing data on
the momentum dependence in the $B\rightarrow D^*$ transition
(the slope
of the Isgur-Wise function) we obtain a new
lower bound on the parameter $\mu_\pi^2 = (2M_B)^{-1}\langle B
|\bar b (i\vec{D})^2 b |B\rangle $ proportional to the
average kinetic energy of $b$ quark inside $B$ meson. The existing
data suggest
$\mu_\pi^2 > 0.4$~GeV$^2$ and (from the ``optical'' sum rule)
$\overline{\Lambda} > 500$ MeV,  albeit with
some  numerical
uncertainties.

\end{titlepage}
\addtocounter{footnote}{-2}

1. In two recent papers \cite{optsr,SUV} it was shown
how the operator product expansion (OPE) allows one to derive
various useful sum rules for
heavy flavor transitions in the small velocity (SV) limit \cite{SV}.
Non-perturbative corrections are included in the theoretical side of
the sum rules
in the form of  an expansion
in the inverse powers of the heavy quark mass. In Ref. \cite{SUV}
the so called first sum rule at zero recoil was obtained which was
then used for estimating the deviation of the $B\rightarrow D^*$
transition form factor from unity
at zero recoil to order ${\cal O}(\Lambda_{\rm
QCD}^2)$. Another sum rule analyzed in \cite{optsr} yields a
field-theoretic proof of the inequality
\begin{equation}
\mu_\pi^2 > \mu_G^2
\label{mumu}
\end{equation}
where $ \mu_\pi^2$ and $ \mu_G^2$ are related to the kinetic
energy and the chromomagnetic operators,
\begin{equation}
\mu_\pi^2 =\frac{1}{2M_B}\langle B|\bar b (i\vec D )^2b|B\rangle ,
\,\,\,
\mu_G^2 =\frac{1}{2M_B}\langle B|\bar b (i/2)\sigma G b|B\rangle .
\end{equation}
(Previously this inequality was obtained within a
quantum-mechanical approach \cite{motion,voloshin}.)  In the
present paper we exploit similar ideas to get a new sum rule in the
SV limit which relates  $ \mu_\pi^2$ to the average square of the
excitation energy of the final hadronic state
$X_c$ in the $B\rightarrow X_c$  semileptonic transitions. At present
the corresponding inclusive differential distribution is not yet
measured.
However, we use the existing data on $B\rightarrow D^* l\nu$
decays near zero recoil to get
 a lower bound on
$\mu_\pi^2$, with no reference to
$\mu_G^2$. The bound involves the slope of the
Isgur-Wise function extracted from
the momentum dependence of the $B\rightarrow D^*$
transition.  Numerically this bound turns out to be close to
 that of Eq. (\ref{mumu}).

\vspace{0.3cm}

2. The general method allowing one to derive the sum rules in the SV
limit is presented in Ref. \cite{optsr}. Here we remind only some
basic points primarily for the purpose of introducing relevant
notations.
Operator product expansion is applied to the transition operator
\cite{1,chay}
\begin{equation}
{\hat T}_{ab}(q) = i\int d^4 x \;{\rm e}^{iqx} \, T\{ j_a^\dagger (x)
j_b (0)\}
\label{1sr}
\end{equation}
where $j_a$ denotes a current of the type
$\bar{c}\Gamma_{a}b$ with an arbitrary Dirac matrix
$\Gamma_a$; $q$ is the momentum carried away by
the lepton pair. The average of ${\hat T}_{ab}$ over
the heavy hadron  state $H_b$ with
momentum $p_{H_b}$  represents a forward scattering
amplitude (the so-called hadronic tensor),
\begin{equation}
h_{ab}(p_{H_b},q) = \frac{1}{2M_{H_b}}
\langle H_b|\hat{T}_{ab} | H_b\rangle\;.
\label{0}
\end{equation}
The observable distributions are expressed through the
structure functions $w_{ab}$,
$$
w_{ab} = (1/i)\, {\rm disc}\,\, h_{ab}.
$$
In the HQET limit \cite{HQET}, when one neglects $1/m_b$, $1/m_c$
corrections, the hadronic tensor $h_{ab}$ is defined by one invariant
function
$h$ for any matrix $\Gamma_a$ in the current $j_a$, namely:
\begin{equation}
h_{ab}~=~C_{ab}h\, , \;\;\;
C_{ab}~=~{\rm Tr} \left[\frac{1 +  \not \!\! v_1}{2}
{\overline\Gamma}_a
 \frac{1 +  {\not \!\! v}_2}{2} \Gamma_b \right]\, .
\label{cab}
\end{equation}
Here
${\overline\Gamma}_a=\gamma_0\Gamma_a^\dagger\gamma_0$
and
$v_{1\mu}$, $v_{2\mu}$ are 4-velocities of
the initial and final  hadrons,
\begin{equation}
v_{1\mu} ~=~\frac{(p_{H_b})_{\mu}}{M_{H_b}}~,\;\;
v_{2\mu} ~=~\frac{(p_{H_b}- q)_{\mu}}{M_{H_c}}~,
\end{equation}
($M_{H_b}$ and $M_{H_c}$ can be substituted by $m_b$ and $m_c$,
correspondingly,  in the leading approximation).
The invariant function
$h$ depends on two scalar invariants
available in the process, namely
$(v_1q)$ and $q^2$. In what follows we will
assume that the hadron $H_b$ is at rest; the first invariant  then
reduces to $q_0$. Moreover,  in  studying the transitions
$b\rightarrow c$ at zero recoil or in the
small velocity (SV) limit,
it is convenient to employ directly the spacelike momentum transfer
$\vec{q}\,^2= (v_1q)^2-q^2$ as the second argument of $h$.

Taking into account higher order in $1/m_{b,c}$ we loose,
generally speaking, this property of the factorization
of $h_{ab}$ into a universal kinematical structure times
one hadronic function $h$.
In the general case, the   hadronic tensor  can be decomposed in
terms
of possible covariants \cite{chay} (their number depends on the
 Lorentz structure
of the currents)
with coefficients  $h_i$. In particular,
in the case of the vector and axial-vector currents
we deal with the functions $h_i^{VV}$,
$h_i^{AA}$ and $h_i^{VA}$, $i=1,...,5$ introduced in
Ref. \cite{koyrakh}. They are
independent functions. However, in the leading order they all are
expressible in terms of $h$, namely,
$$
h=\frac{h_1^{AA}}{1+v_1v_2} =
\frac{1}{2}\frac{m_c}{m_b}h_2^{AA} = -m_c h_5^{AA} ;
$$

$$
h=\frac{h_1^{VV}}{1-v_1v_2} =
-\frac{1}{2}\frac{m_c}{m_b}h_2^{VV} = m_c h_5^{VV} ;
$$
and
$$
h= -m_c h_3^{VA}.
$$
The functions $h_i$ not listed here vanish in this approximation.
The expressions for all functions $h_i$  up to order
$1/m_{b}^2$ can be found in \cite{koyrakh}.

Although the universal factorization above is not valid
for all non-perturbative corrections it still holds for
those corrections that are relevant for the third sum rule
to be derived below.  We will explain this point shortly.
Since it is not important what hadronic function we deal
with -- they all lead to one and the same third sum rule --
we will use $h_1^{AA}$ in our derivation.
Thus, we consider the transitions of the $B$ meson
induced by the axial-vector current,
$$
A_\mu = \bar c\gamma_\mu\gamma_5 b\; .
$$
To single out $h_1^{AA}$ one must consider the spatial
components of the axial current generating the transitions of the
$B$ meson to $D^*$ and the corresponding excitations.

In \cite{SUV}
the
sum rules at zero recoil ($\vec q =0$) were obtained; here we will
work at small but non-vanishing values of $|\vec q |$. The terms
${\cal O}({\vec q}^2)$ will be kept while those of higher order in
$|\vec q |$ will be neglected.

To start the derivation we consider $h_1^{AA}(q_0, \vec q )$ in the
complex
$q_0$ plane  ($\vec q$ is assumed to be fixed, and $\Lambda_{\rm
QCD}
\ll |\vec  q
|\ll M_D$). More exactly, let us shift $q_0$ by introducing the
quantity
\begin{equation}
\epsilon = q_{0max}-q_0
\end{equation}
where
\begin{equation}
q_{0max} = M_B- E_{D^*},\,\,\, E_{D^*}= M_{D^*}+\frac{{\vec
q}^2}{2M_{D^*}}.
\label{q0max}
\end{equation}
When $\epsilon$ is real and positive we are on the cut. The
imaginary part of $h_1$ is given by the ``elastic" contribution of
$D^*$ plus inelastic  excitations. For what follows it is crucial that all
these contributions are
{\em positive definite}.

For negative $\epsilon$ we are below the cut, and the amplitude
$h_1^{AA}$ can be computed -- and it actually was
\cite{koyrakh,k2,k3} -- as an expansion in $\Lambda_{\rm
QCD}/m_{c,b}$. For our purposes it is sufficient to limit ourselves to
the correction terms of the first and the second order in
$\Lambda_{\rm QCD}$. This is exactly the approximation adopted in
\cite{koyrakh,k2,k3}, and  expressions obtained there will be used
below.

At the next stage we  assume that $\Lambda_{\rm QCD}
\ll |\epsilon |\ll  m_{b,c}$. The amplitude $h_1^{AA}$ is expanded
in powers of $\Lambda_{\rm QCD}/\epsilon$ and
$\epsilon /m_{b,c}$. Polynomials in $\epsilon$ can be discarded since
they have no imaginary part. We are interested only in negative
powers of $\epsilon$. The coefficients in front of $1/\epsilon^n$
are related, through dispersion relations, to the integrals over the
imaginary part of $h_1^{AA}$ with the weight functions proportional
to the excitation energy to the power $n-1$. Thus, the first sum rule
considered in Ref. \cite{SUV} corresponds to $n=1$; the second sum
rule (sometimes called optical or Voloshin's sum rule \cite{volopt},
see also \cite{Grozin,burk}) corresponds to $n=2$. The lower bound
on
$\mu_\pi^2$ -- our main aim in this work -- stems from the third
sum rule, i.e. we need to analyze the coefficient in front of
$1/\epsilon^3$ in the expansion of $h_1$.

The $1/\epsilon$ expansion can be read off from Eq. (A.1) in Ref.
\cite{koyrakh}.  One technical element of the derivation deserves a
comment. The theoretical expression for the amplitude $h_1^{AA}$
presented in \cite{koyrakh} knows nothing, of course, about the
meson masses; it contains only the quark masses. Correspondingly,
it is convenient to built first the expansion of $h_1^{AA}$ in an
auxiliary quantity,
\begin{equation}
\epsilon_q = m_b- E_c - q_0,\,\,\, E_c =m_c +\frac{{\vec q}^2}{2m_c}.
\end{equation}
Then, if necessary, we  reexpress the expansion obtained in this way
in terms of $\epsilon$. The difference between
$\epsilon_q$ and $\epsilon$ is ${\cal O}(\Lambda_{\rm
QCD}\cdot{\vec
q}^2/m_{b,c}^2)$ and ${\cal O}(\Lambda_{\rm QCD}^2/m_{b,c})$. It
will
be seen shortly that to our accuracy  this difference can be simply
ignored in the  third sum rule in the SV limit. It can not be
discarded, however, in  the second sum rule. (The situation is quite
different from that which  took place in the sum rules at zero recoil,
see \cite{optsr}. At zero  recoil the difference between $\epsilon$
and $\epsilon_q$ was  absolutely important.)

The expression for $h_1^{AA}$ in Eq. (A.1) in \cite{koyrakh}
has the form
\begin{equation}
-h_1^{AA}
=[(m_b+m_c-q_0) +{\cal O}(\Lambda_{\rm QCD}^2/m_b)]
\frac{1}{z}
+ {\cal O}(\Lambda_{\rm QCD}^2) \frac{1}{z^2}+
 \frac{4}{3}(m_b+m_c-q_0)\mu_\pi^2{\vec q}^2\frac{1}{z^3}
\end{equation}
 where
\begin{equation}
z=\epsilon_q (2E_c +\epsilon_q ).
\end{equation}
Notice the similarity of the coefficient in front of $1/z^3$
and the leading part of the coefficient in front of $1/z$.
This is not accidental. The terms $1/z^3$
appear only as the expansion of the second order in $\pi q$
of the denominator
$
(m_b v - q)^2 -2q\pi
$
(see Ref. \cite{koyrakh}) and, therefore, preserve
the same universal factorization which was pointed out
above in the HQET limit.

Expanding in $\epsilon_q/2E_c$ we observe that
$1/z^n$ reduces to $1/\epsilon_q^n$ plus all lower
powers of $1/\epsilon_q$ plus a polynomial  in
$\epsilon_q$.
The next step is eliminating  $\epsilon_q$
in favor of $\epsilon$. The term $1/z^3$ comes
with a  coefficient $\mu_\pi^2\cdot{\vec q}^2$; hence here the
difference between $1/\epsilon$ and $1/\epsilon_q$ is of higher
order and can be neglected.
By the same token to order ${\cal O}(\Lambda_{\rm QCD}^2)$ one can
substitute $1/\epsilon_q$ by $1/\epsilon$ in $1/z^2$. As far as
$1/z$ is concerned here we must reexpress $1/\epsilon_q$ in terms
of $1/\epsilon$,
\begin{equation}
\frac{1}{\epsilon_q} = \frac{1}{\epsilon} + \frac{(\epsilon
-\epsilon_q)}{\epsilon^2} +...
\label{epsq}
\end{equation}
Next terms in Eq. (\ref{epsq}) are irrelevant since they lead to
corrections of higher order in $\Lambda_{\rm QCD}$ and/or
$|\vec q |$. This observation is crucial since it tells us that the
$1/z$ part contributes only to the first and the second sum rules;
it generates no $1/\epsilon^3$ terms. As a result
$h_1^{AA}$ has the form
$$
-h_1^{AA}= \frac{1}{\epsilon}
\left( 1 -\frac{{\vec q}^2}{4m_c^2} +{\cal O}(\Lambda_{\rm
QCD}^2/m_c^2)\right) +\frac{1}{\epsilon^2}
\left({\cal O}(\Lambda_{\rm QCD}^2/m_c)+ {\cal O}(\Lambda_{\rm
QCD}{\vec q}^2/m_c^2)
\right)+
$$
\begin{equation}
\frac{1}{\epsilon^3}
\frac{\mu_\pi^2}{3}  \, \frac{{\vec q}^2}{m_c^2} \,
+{\rm polynomial}\, .
\label{h1}
\end{equation}
Here only the terms ${\cal O}({\vec q}^2)$ are kept. We also
do not discuss perturbative corrections. Writing out the
dispersion relation in $\epsilon$,
$$
-h_1^{AA} (\epsilon ,{\vec q}^2) =
\frac{1}{2\pi}\int d\tilde\epsilon\,\frac{w_1^{AA}(
\tilde\epsilon ,{\vec q}^2)}{\epsilon -\tilde\epsilon}=
$$
\begin{equation}
\frac{1}{\epsilon}\cdot\frac{1}{2\pi}\int
d\tilde\epsilon\,w_1^{AA}(
\tilde\epsilon ,{\vec q}^2)+
\frac{1}{\epsilon^2}\cdot\frac{1}{2\pi}\int
d\tilde\epsilon\,\tilde\epsilon\, w_1^{AA}(
\tilde\epsilon ,{\vec q}^2)+
\frac{1}{\epsilon^3}\cdot\frac{1}{2\pi}\int
d\tilde\epsilon\,{\tilde\epsilon}^2 w_1^{AA}(
\tilde\epsilon ,{\vec q}^2)+...
\label{disp}
\end{equation}
and expanding it in $1/\epsilon$ we get
 the sum rules
 by equating the coefficients in front of $1/\epsilon^n$. Here
$w_1^{AA} = 2 \, {\rm Im} \, h_1^{AA}$.

\vspace{0.3cm}

3. Now let us discuss the phenomenological side of the sum rule.
The structure function
$w_1^{AA}$ is non-vanishing for positive $\epsilon$,
\begin{equation}
w_1^{AA} (\epsilon) =
\sum_{i=0}^{\infty}\frac{|F_{B\rightarrow i}|^2}{2E_i} \, 2\pi
\delta (\epsilon
- \delta_i ) ,
\label{sum}
\end{equation}
where the sum runs over all possible final hadronic states,
the term with $i=0$ corresponds to the ``elastic" transition
$B\rightarrow D^*$ while $i=1,2,\ldots$ represent excited states
with the energies $E_i = M_i +{\vec q}^2/(2M_i)$.
Strictly speaking, $|F_{B\rightarrow i}|^2$ does not present
the square of a form factor; rather this is the contribution to the
given structure function coming from the multiplet of the
degenerate states which includes summation over spin states as well.
In the particular example considered $D$ is not produced
in the elastic transition, so that in the elastic part one needs
to sum only over polarization of $D^*$. Therefore, the term
``form factor" for  $F_{B\rightarrow i}$ is rather symbolic.
$|F_{B\rightarrow i}|^2$  depends on  $\vec q$.
Moreover, $\delta_i$ in Eq. (\ref{sum}) is the excitation energy
(including the corresponding kinetic energy),
$$
\delta_i = E_i - E_{D^*} .
$$
For the elastic transition $\delta_0$ vanishes, of course.

 The dispersion representation (\ref{disp}) and Eq. (\ref{h1})
 lead to the following sum rule for the second moment
of $w_1^{AA}$ (the coefficient in front of
 $1/\epsilon^3$, the third sum rule in the nomenclature of Ref.
\cite{optsr}):
\begin{equation}
\frac{1}{2\pi}\int d\epsilon \epsilon^2w_1^{AA}(\epsilon ) =
\sum_{i=1}^{\infty}\frac{|F_{B\rightarrow i}|^2}{2E_i}
\delta_i^2 = \frac{1}{3}\mu_\pi^2\frac{{\vec q}^2}{m_c^2}.
\label{sr}
\end{equation}

We pause here to make a few remarks regarding Eq. (\ref{sr}).
First of all, since $\delta_0 =0$, this kills the elastic contribution in
the left-hand side, and the sum actually starts from the first
excitation. Second, since all $\delta_i^2$ are of order $\Lambda_{\rm
QCD}^2$ we need to know $F_{B\rightarrow 2}$, $F_{B\rightarrow
3}$, etc. only to the zero order in $\Lambda_{\rm QCD}$. To this
order all transition form factors to the excited states are
proportional to $\vec q$, i.e.
\begin{equation}
|F_{B\rightarrow i}|^2\propto {\vec q}^2 .
\label{F}
\end{equation}
(As a matter of fact,
 the transitions to $P$-wave states are relevant, see \cite{IW} for
further details.) Moreover, taking account of Eq. (\ref{F})
we can neglect ${\cal O}({\vec q}^2)$ part in $\delta_i$'s, so that
in Eq. (\ref{sr})
$$
\delta_i = M_i -M_{D^*} .
$$
Third, $m_c^{-2}$ in the right-hand side can be replaced, to the
accuracy desired,  by $(M_{D^*})^{-2}$ or by the mass of any excited
state.

After all these simplifications the third sum rule takes the form
\begin{equation}
\sum_{i=1}^{\infty}\frac{|F_{B\rightarrow i}|^2}{2M_i}
(M_i-M_{D^*})^2 = \frac{1}{3}\mu_\pi^2{\vec v}^2,
\label{sum3}
\end{equation}
where
$$
\vec v =\frac{\vec q}{M}
$$
(it does not matter whose particular mass, $M_{D^*}$ or $M_i$,
stands in the denominator).

The next steps are rather obvious. The lower bound on $\mu_\pi^2$
is a consequence of positivity of all individual contributions in the
left-hand side of Eq. (\ref{sum3}). Indeed, let us rewrite it as
follows:
\begin{equation}
\frac{1}{3}\mu_\pi^2 = \delta_1^2\cdot\sum_{i=1}^{\infty}
\frac{|F_{B\rightarrow i}|^2}{2M_i{\vec v}^2} +
\sum_{i=2}^{\infty}\frac{|F_{B\rightarrow i}|^2}{2M_i{\vec
v}^2}(\delta_i^2-
\delta_1^2).
\label{mu3}
\end{equation}
The second term is evidently positive. The first sum
can be found, in turn, by using the
Bjorken sum rule \cite{bj}. This sum rule relates  the sum
over the $P$-wave states in
the brackets
 to the ${\vec q}^2$ dependence of the ``elastic" $B\rightarrow
D^*$ transition (the
slope of the Isgur-Wise function \cite{IW2}).

\vspace{0.3cm}

4. It is instructive to briefly reiterate derivation of the Bjorken sum
rule, which, as explained above, is needed only in the zero order in
$\Lambda_{\rm QCD}$.
Equating the coefficients of $1/\epsilon$ in Eqs. (\ref{h1})
and (\ref{disp}) one immediately finds
\begin{equation}
\frac{1}{2\pi}\int d\epsilon \; w_1^{AA}(\epsilon ) =
\sum_{i=0}^{\infty} \frac{|F_{B\rightarrow i}|^2}{2E_i} =  1-
\frac{{\vec
v}^2}{4}\, .
\label{bjorken}
\end{equation}
The elastic part here can be parametrized in terms of the Isgur-Wise
function $\xi (v_1v_2)$ \cite{IW2,falk}. The $B\rightarrow D^*$
transition has  the
form
$$
\langle D^*(v_2)|A_\mu |B(v_1)\rangle = \sqrt{M_BM_{D^*}} \,
\left[ \epsilon_\mu (1+v_1v_2) -(\epsilon v_1)v_{2\mu}\right] \xi
(v_1v_2)
$$
where $v_{1,2}$ are the four-velocities. This means that
\begin{equation}
(2E_{D^*})^{-1}|F_{B\rightarrow D^*}|^2= \frac{M_{D^*}}{E_{D^*}}
\left(
\frac{1+v_1v_2}{2}
\right)^2 |\xi (v_1v_2)|^2 \approx
1 - \rho^2 {\vec v}^2\, .
\label{elastic}
\end{equation}
Here $\rho^2$ is the slope parameter \cite{bj},
\begin{equation}
\xi (v_1v_2) = 1 - \rho^2 (v_1v_2-1) +... = 1-\rho^2\frac{{\vec
v}^2}{2} +...
\end{equation}
and we used the fact that $\xi$ at zero recoil is unity \cite{SV}.

Notice that although we discuss the Bjorken sum rule
for the axial current actually it can be derived for
arbitrary current $j_a =\bar b \Gamma_a c $. To the leading order in
$1/m_{b,c}$ the universal factorization (\ref{cab}) takes
place for the structure functions $w_{ab} = 2{\rm Im}\, h_{ab}$.
Moreover, the sum over any HQET degenerate multiplet of states
gives
\begin{equation}
\frac{1}{2M_B}\sum_i \langle B|j_a^{\dagger}|H_c^i \rangle
\langle H_c^i|j_b|B \rangle ~=~ C_{ab}M_{H_c}\frac{1+v_1 v_2}{2}
|\xi_{H_c}(v_1 v_2)|^2
\end{equation}
where $\xi_{H_c}$ is the Isgur-Wise function for the $H_c$ multiplet.

At $\vec v =0$ the sum rule (\ref{bjorken}) is trivially satisfied
since at zero recoil all inelastic form factors vanish, and we are left
with the elastic contribution which reduces to unity. The term
linear in ${\vec v }^2$ yields a relation between the slope of $\xi$
and the inelastic contributions,
\begin{equation}
\rho^2-\frac{1}{4}= \sum_{i=1}^{\infty}
 \frac{|F_{B\rightarrow i}|^2}{2M_i{\vec v}^2}
\label{rhobj}
\end{equation}
Let us remind that the ratio $|F_{B\rightarrow i}|^2/{\vec v}^{2}$
has the finite limit at zero recoil.
 Eq. (\ref{rhobj}) is the Bjorken sum rule proper
\cite{bj}. Let us add for completeness that in the notations of Ref.
\cite{IW}, where the $P$ wave
inelastic contributions are written out explicitly, it takes the
form
$$
\rho^2-\frac{1}{4} =\sum_{n=1}^\infty
|\tau_{1/2}^{(n)}(1)|^2 +2\sum_{n=1}^\infty
|\tau_{3/2}^{(n)}(1) |^2
$$
(for a simple derivation see Ref. \cite{Grozin}).
{}From these expressions it follows, in particular, that $\rho^2>1/4$.

Combining Eq. (\ref{rhobj}) with Eq. (\ref{mu3}) we finally arrive at
\begin{equation}
\mu_\pi^2 = 3\delta_1^2\left(\rho^2 -\frac{1}{4}\right) +
3\sum_{i=2}^{\infty} \frac{|F_{B\rightarrow i}|^2}{2M_i{\vec
v}^2}(\delta_i^2-
\delta_1^2)\, ,
\label{mu}
\end{equation}
$$
\delta_i = M_i -M_{D^*}\, .
$$

Eq. (\ref{mu}) is  a direct $n=3$ generalization of
Voloshin's sum rule written for $n=2$ \cite{volopt}, see also
\cite{optsr},
\begin{equation}
\overline{\Lambda}= 2\delta_1\, (\rho^2-\frac{1}{4})   +
2\sum_{i= 2}^\infty
 \frac{|F_{B\rightarrow i}|^2}{2M_i{\vec v}^2}(\delta_i-
\delta_1)
\, .
\label{11v}
\end{equation}

Since the second term in Eq. (\ref{mu}) is positive we get
the following obvious inequality:
\begin{equation}
\mu_\pi^2 >  3\delta_1^2\, (\rho^2-\frac{1}{4})
\label{12}
\end{equation}
(we remind that $\delta_1$ here is the lowest excitation energy,
$\delta_1 = M_1-M_{D^*}$).  For a
first, rough,
estimate let us assume that $\delta_1 \approx 500$ MeV and use the
central
value of the measured slope \cite{Cassel} of the $B\rightarrow D^*$
form
factor for $\rho^2$ ,
\begin{equation}
\rho^2 =
0.84\pm 0.12\pm 0.08 .
\end{equation}
 Then
$$
\mu_\pi^2> 0.45 \,\, {\rm GeV}^2.
$$

Three comments are in order here regarding the sum rules presented
above.  First, the very same final results are obtained irrespectively
of what currents we start from, axial or vector, or a mixture of these
two. The only difference is that, say, for the vector currents we
would
get $M_D$, not $M_{D^*}$ in the definition of $\delta_1$.  This
difference is unimportant in the limit $m_{b,c}\rightarrow\infty$, of
course. This remark brings us to the second point.
In Eq. (\ref{12}) all subleading $1/m_{b,c}$ terms have been
omitted; thus, all quantities there
refer to the infinite mass (static) limit, and the corresponding
hadronic
parameters must
be understood just in this sense. In other words, rather than using
the experimental value of $\rho^2$ and $\delta_1$ measured in the
beauty-to-charm transitions and the charmed family, respectively,
one should use the static values of $\rho^2$ and $\delta_1$. (Hence,
$M_D$ is indeed equal to $M_{D^*}$ with our accuracy.) Finally, in the
original sum
rules the
sum runs over all states including those which represent
high-energy
excitations described, in the dual sense,  by perturbative formulae
(see Ref. \cite{optsr} for more details). To get predictions for
$\mu_\pi^2$ and
$\overline{\Lambda}$
normalized at a low (quark-mass independent) scale $\mu$
 one must  truncate the sum over the excited states at
$\delta_i\sim \mu$ and invoke
 duality  between the perturbative corrections and  the contributions
of the excited states above $\mu$.

In general, the sum rules at non-zero recoil get $\Lambda_{\rm
QCD}/m_{b,c}$ corrections
which depend on the particular choice of the weak
current  considered and
can be sizable. However, all corrections
to the hadronic tensor $h_{ab}$
start with terms
explicitly proportional to
$\Lambda_{\rm QCD}^2/m_{b,c}^2$ \cite{chay,6,6a}, see Eq. (A.1) in
Ref. \cite{koyrakh}. The question is where the linear corrections
come from? A source of subleading corrections is quite obvious: they
appear at the stage
when one expresses $\epsilon_q$ in the theoretical formulae in
terms of $\epsilon$; since $M_B =m_b +\overline{\Lambda}+...$ (and
the same for the
charmed quark) they contain linear terms.
This does not affect, of course,  the first sum rule ($n=1$), and in
this case the prediction starts from unity plus  corrections
at the   level $\overline{\Lambda}^2/m_{b,c}^2$ \cite{SUV}.

\vspace{0.3cm}

5. We proceed now to a more careful discussion of the numerical
situation.
The experimentally measured $B\rightarrow
D^*(\mbox{unpolarized})\;l\nu$
decay rate is expressed in terms of the Isgur-Wise function in
the leading approximation, see Eq. (\ref{elastic}). In this
approximation the slope of the Isgur-Wise function is related
to the ${\vec q}^2$ dependence of the $B\rightarrow D^*$ rate. It
 is clear that with
$1/m_{b,c}$ and radiative corrections included the ${\vec q}^2$
dependence of the decay rate does not exactly coincide any more
with
the slope of the Isgur-Wise function.
These corrections were  estimated in the literature (see the
review paper \cite{nrev}). Their effect seems to be
equivalent to a  decrease in
the  slope of  the Isgur-Wise function,
 by about
9\% .  At the same time the radiative perturbative
corrections were found~\footnote{Although the sign of the
perturbative
effects is obvious on general physical grounds, we think that the
concrete procedure of evaluating them described in Ref. \cite{nrev}
systematically
overestimates the velocity dependence, at least for the observable
we discuss
here.} to  increase the  slope by $\sim$ 20\%.
Therefore, taking these estimates at their face value  we
are inclined to conclude that
\begin{equation}
\rho^2 \approx {\rho}^2_{exp} - 0.1
\label{15}
\end{equation}
where ${\rho}^2$ in the left-hand side is the genuine static value of
the slope parameter while ${\rho}^2_{exp}$ is the value of the
parameter obtained from
experimental fits
of the dependence of the $B\rightarrow
D^*(\mbox{unpolarized})\;l\nu$
decay rate on the $D^*$ velocity. It is worth emphasizing again that
the slope $\rho^2$ of the Isgur-Wise function by
definition does not depend on the structure of weak currents
considered. The
above numerical estimates of both, perturbative and $1/m_{b,c}$
corrections, have
been obtained for the real $V-A$ current to which experimental
numbers refer.

Similar effects due to the finite mass of the
$c$ quark enter our lower bound implicitly when we use
the observed mass values of the
excited charmed mesons.
In the future these pre-asymptotic corrections can be
isolated in a model-independent way once the
masses of the beauty counterparts are measured. The most
sizable
corrections are expected due to the chromomagnetic interaction
of the heavy quark
spin inducing hyperfine splitting among the members of the heavy
spin
multiplets. In
particular,  $M_{D^*}-M_D \sim 140$ MeV. This effect is presumably
accounted
for by substituting the spin averaged masses for the ground
$S$-wave states and for the $P$-wave excitations, rather than actual
masses of $D$, $D^*$, etc. We actually did this spin averaging.
Another shift arises due to
the heavy
quark kinetic
energy term in the hadron mass. It is natural to expect its value to
be
smaller in the excited mesons than for the ground state. Therefore,
the
static limit  of $\delta_1$ is  expected  to be somewhat
larger than the value of $\delta_1$
 experimentally
observed for the actual charmed particles, but probably  not more
than by 50 MeV.
 We then
use  the value
\begin{equation}
\delta_1 \approx 500 \,\, {\rm MeV}
\label{17}
\end{equation}
as a very reasonable educated guess for the static value of
$\delta_1$.

With the parameters from  Eqs. (\ref{15}), (\ref{17})
we finally get
$$
\mu_\pi^2 > 0.37 \,\, {\rm GeV}^2\; ,
$$
\begin{equation}
\overline{\Lambda} >500\,\, {\rm MeV}
\label{18}
\end{equation}
where the second relation comes from Voloshin's sum rule
(\ref{11v}). These lower bounds are seen to lie not very far from the
estimates obtained earlier within  QCD sum rules \cite{Braun}
\begin{equation}
\mu_\pi^2\sim 0.55\,\, {\rm GeV}^2\; ,\;\;\;\;\;\;\overline{\Lambda}
\sim
450\,\,{\rm MeV}\; .
\label{19}
\end{equation}
Note that the lower bound on $\mu_\pi^2$ in Eq. (\ref{18})
is numerically quite close
to the bound (\ref{mumu}) derived recently in
\cite{motion,voloshin}.

Unfortunately, numerical uncertainties in all the numbers
above prevent us from making too strong a statement. Nevertheless,
let us assume for a moment that
future
refined measurements and calculations of the subleading corrections
in the third sum rule will confirm these values and
establish the fact
that  two
inequalities in Eq. (\ref{18}) are rather close to saturation. This
would
mean that
the
sum rules are actually saturated -- to a reasonable
degree of accuracy -- by the contributions from  the
states with masses around $M_D+\delta_1$ generically
called
$D^{**}$ in this context. To account for nonperturbative effects in
$b\rightarrow c$
decays  one then would need only to consider one inelastic channel,
``$D^{**}$''. The higher excited states will be  represented (in the
sense of duality) by purely perturbative probabilities calculated
in the
free quark-gluon approximation. We actually consider such a
situation as a most  natural scenario in QCD.
It is worth noting that the $D\pi$ contribution to the third sum rule
is suppressed for soft pions, unlike the first sum rule, where it was
quite substantial \cite{SUV}. The effect of the ``hard" pion emission
is well represented by some of the $P$-wave $D^{**}$ resonances.

\vspace{0.3cm}

6. We have derived the third sum rule for the $b\rightarrow c$
transition in the SV limit and showed how one can use it to constrain
the kinetic energy parameter $\mu_\pi^2$ by using the data on
$B\rightarrow D^*$. In principle it is quite conceivable that the full
differential distribution in $q_0$ and ${\vec q}^2$ in the
inclusive semileptonic $B$ decays
will be measured in  the future.
This  measurement can then be  immediately
translated in the value of $\mu_\pi^2$, one of the most important
parameters of the heavy quark physics.
 The more one  will learn about the decays to the excited
states the more accurate the determination of $\mu_\pi^2$ will
become.

\vspace*{.3cm}

{\bf ACKNOWLEDGMENTS:} \hspace{.4em} We are grateful
to
L.~Bellantoni, D.~Cassel, Y.~Kubota and R.~Poling  for useful
discussions of  the current experimental situation. A.G. and N.U.
would like to thanks the  CERN  Theory Division  for hospitality.
This work was supported in part by the National Science Foundation
under the grant number
PHY 92-13313, and by DOE under the grant number
 DE-FG02-94ER40823,  and by Royal Society and PPARC, UK.


\end{document}